# Nonlinear digital imaging

Jen-Tang Lu, Alexandre Goy, and Jason W. Fleischer*

Department of Electrical Engineering, Princeton University

*e-mail: jasonf@princeton.edu

**Abstract**:

Nonlinear imaging systems can surpass the limits of linear optics, but to date they have all relied on physical media (e.g. crystals) to work. These materials are all constrained by their physical properties, such as frequency selectivity, environmental sensitivity, time behavior, and fixed nonlinear response. Here, we show that electro-optic spatial light modulators (SLMs) can take the place of traditional nonlinear media, provided that there is a feedback between the shape of the object and the pattern on the modulator. This feedback creates a designer illumination that generalizes the field of adaptive optics to include object-dependent patterns. Unlike physical media, the SLM response can provide arbitrary mathematical functions, operate over broad bandwidths at high speeds, and work equally well at high power and single-photon levels. We demonstrate the method experimentally for both coherent and incoherent light.

Illumination methods in imaging have lagged behind other advances in computational optics. In most systems, the illumination has a constant intensity, so that the object of interest is sampled uniformly. Examples range from sunlight and studio light in everyday photography to Kohler illumination in microscopy and CAT scans in biomedicine. Structured illumination, in which the incident intensity is patterned, provides more information at the input of the system and can be used to extract more information at the output. Perhaps the earliest and most dramatic example is the Zernike phase contrast technique [1], which converts phase objects into intensity variations suitable for direct measurement. Phase contrast is implemented by using a particular illumination pattern (typically a ring in the Fourier plane) and a phase filter in a conjugate plane that matches the shape of the illumination. Other examples of structured illumination include periodic masks, which can lead to improved resolution [2, 3] and depth sectioning [4], Bessel [5] and Airy [6] beams, which lead to extended depth-of-focus, random masks, which enable single-pixel cameras [7], and coherence engineering [8, 9], with which the statistics of the lighting can be controlled. All these techniques use illumination patterns that are independent of the sample. Primarily, this is because the object is typically unknown and a universal form of illumination is desired. This creates a general, shift-invariant transfer function that is suitable for arbitrary objects but is optimized for none of them.

In parallel with structured illumination, adaptive optics techniques have been established in which scattering and aberrations along an optical path are "pre-compensated" by first measuring and then correcting for the distortions, e.g. using deformable mirror arrays. This has enabled a revolution in remote sensing and ground-based astronomy, e.g. for imaging through a varying atmosphere, and is used routinely in ophthalmology for optical examination of the retina. It is also experiencing resurgence in imaging through turbid media, such as paint and fog [10, 11], in which scatterers randomize the information from object to image. As before, however, the adaptation is independent of the object, used instead for correcting the path along the rest of the optical system.

Here, we introduce a new type of adaptive optics in which the illumination of the object depends on the shape of the object itself. This gives a feedback loop between the (digital) image of the object and the (digital) pattern of illumination, effectively creating a nonlinearity that can be controlled by an electro-optic modulator. Unlike nonlinearities that are determined – and limited – by physical media, such as crystals and polymers, spatial light modulators work for any incident intensity, operate over broad bandwidths at high speeds, work for any degree of coherence, and can have nearly arbitrary mathematical form. Further, the digital nonlinearities can improve traditional imaging figures of merit, such as resolution and contrast [12], and can be tuned for task-based imaging [13], such as correlation and edge detection.

The experimental design is shown in Fig. 1A. The method consists of two basic steps. First, a standard image is obtained using uniform illumination. Second, the measured pattern is fed into a spatial light modulator and used to re-illuminate the object. While the functional relationship between the SLM and the object is essentially arbitrary, we consider here the simplest case of an intensity-dependent phase modulation: $\phi_{SLM} = \gamma I^\alpha$, where $\gamma$ is the strength of the modulation and α is its order. This makes the system a physical implementation of the conventional split-step method in beam propagation codes, with one step for diffraction and one step for nonlinear effect, and allows a straightforward comparison with more conventional nonlinearities, e.g. the Kerr response for which α = 1. We emphasize, though, that much more complex responses are possible, including the ability to impose separate amplitude and phase modulations using a second SLM. More details about the digital nonlinear response are given in the Supplementary Information.

We experimentally demonstrate the method by nonlinearly improving the imaging of a phase object (Fig. 1C). Here, we use a modified version of the Gerchberg-Saxton method [14] to retrieve the phase, taking advantage of the fact that phase matching from the nonlinear illumination-object feedback provides a strict constraint on the behavior of the algorithm [15]. As the shape of the object is not known *a priori*, we first record a defocused intensity ($I_U=|A_U|^2$) that is obtained using uniform illumination (Fig. 1D). The amplitude distribution, $|A_U|$, is then fed into a SLM as a phase modulation $\phi_{SLM} \propto |A_U|$, enabling a second intensity measurement ($I_D=|A_D|^2$) with designer illumination (Fig. 1E). As in previous examples of nonlinear imaging [15,16], we use a self-defocusing nonlinearity; this enables sufficient wave mixing to enhance imaging while suppressing the fast instabilities common to self-focusing systems [17, 18].

From the two images $I_U$ and $I_D$, the nonlinear phase retrieval algorithm proceeds in a manner similar to the Gerchberg-Saxton algorithm. First, a field with amplitude $A_U$ and initial uniform phase $\phi = 0$ is numerically back-propagated from the (out-of-focus) camera plane to the sample plane (in focus). The phase modulation $\phi_{SLM}$ is then added to the phase of the field in order to simulate the action of the SLM, and the new field is forward-propagated to the camera plane. This simulated amplitude is then replaced by the amplitude of the actual measurement. Finally, the process is repeated until convergence. A pseudocode for this algorithm is shown in Fig.1B.

As a phase object, we used a sample consisting of a glass plate with the three characters "DIC" etched on it. The in-focus image of the sample is shown in Fig. 1C. As the object is thin and transparent, no intensity modulation appear on the image in focus apart from scattering from the character edges. In Fig. 1D and 1E, we show the measured diffraction intensity distributions in the detector plane (out-of-focus) with uniform and designer (feedback) illumination respectively. In general, more diffraction fringes are visible in the latter image, since the illumination has the shape of the object itself. The second pass of illumination effectively doubles the diffraction pattern; to next order, then, we would expect at least a two-fold increase in sensitivity to the object, leading to double the image quality in the final measurement. Larger multiplication is of course possible *[16],* especially with cascaded interactions [19], with potentially exponential improvement upon repeated iteration.

We compare our nonlinear algorithm with a variation of the linear Gerchberg-Saxton algorithm for the same defocused distance. The linear algorithm relies on one in-focus and one out-of-focus intensity measurement using uniform illumination. Simulation results are shown in Fig. 2A and 2B. In real imaging cases, the target is unknown and reconstructed phase error cannot be accessed; in the simulation test case, the ground truth is chosen *a priori,* and we can quantitatively measure reconstructed phase error

$$\text{ER} = \frac{\sum_{r \epsilon S} |f(r) - g[r]|}{\sum_{r \epsilon S} |g[r]|} \tag{1}$$

where f(r), g(r), and S are the reconstructed image, the ground truth, and the image space, respectively.

As expected, the nonlinear reconstruction (Fig. 2B) is about 2x better than the linear reconstruction (Fig. 2A) in term of the phase error (ER 0.8 vs. 1.5). The experimental counterparts, shown in Figs. 2C-E, confirm the result that the nonlinear algorithm provides much better reconstruction than the linear algorithm. For further comparison, we consider contrast and resolution, defined respectively as

$$Contrast = \bar{I}_{DIC} - \bar{I}_{BKG} \qquad (2)$$

$$R = (\nabla_x I)^{-1}_{x \in DICE} \qquad (3)$$

where $\bar{I}$ represents the mean value of the intensity, the subscript "DIC" denotes the region of the DIC characters, "BKG" denotes the rest of the image, and "DICE" denotes the edge domain of the DIC region. (For a consistent measure of resolution based on a single image, we take the highest resolvable spatial frequency, given by the inverse of the normalized image gradient averaged over all the character edges.) By these metrics, the nonlinear algorithm yields a 260% improvement in contrast (1.50 vs. 0.58) and 115% improvement in resolution (14.8 vs 17.1 microns) over linear reconstruction.

The quality of the reconstructed phase image using the nonlinear algorithm is comparable to what we obtained in a previous work using a nonlinear photorefractive crystal [15]. In that system, the intensity pattern of the object induced an index change in the crystal, creating a self-reinforcing diffraction grating during light propagation. Here, the object-dependent pattern on the spatial light modulator creates a similar grating

(computer-generated hologram) in the plane of the SLM, which acts more like an iterative map. In both cases, the nonlinear feedback means that the grating structure is automatically phase-matched with the object, guaranteeing an optimized diffraction pattern. For the computational part, the interplay between intensity and phase gives an extra constraint on the algorithm, resulting in improved phase sensitivity, selectivity, and convergence.

Another advantage of the digital system, and the corresponding algorithm, is the ability to work for nearly any degree of spatial coherence. We demonstrate this experimentally by making the illumination beam partially spatially coherent by passing it through a rotating diffuser [20]. The performance of linear and nonlinear modulation, in terms of contrast and resolution, are shown in Fig. 3J and 3K, respectively. In all cases, the figures of merit for linear propagation get worse monotonically as the degree of coherence is decreased (consistent with a progressive decrease in visibility of diffraction fringes [21]). In the nonlinear case, both contrast and resolution are improved simultaneously beyond their linear limits. (This win-win situation, normally an engineering trade-off in linear systems [12], is a general feature of nonlinear imaging systems [16]). For designer illumination, the contrast actually improves with decreasing coherence, up to an optimal value, while the system resolution is relatively constant above and below this value. The former is a stochastic resonance effect [22], peaking when the smallest significant feature of the object (the stroke width of the DIC characters, ~ 167μm) matches the spatial correlation length of the illumination (characterized by the speckle spatial frequency $3.75 \times 10^4$ rad/m). The latter results from improved visibility due to mean-field (DC) scattering [16].

Theoretically, the improvements in image quality are unlimited [3, 16]. In practice, there are many factors that hinder the performance. These include the usual trade-off between pixel size and dynamic range in the camera and SLM [23], pixel-mapping and tone-mapping mismatches between devices, amplitude corruption in the phase modulation, and genuine issues of noise

The best approach to retrieving (estimating) the phase object is to maximize the information in the encoded illumination. This is given by the most widespread sampling in phase space, which for the experimental setup here corresponds to the most diverse distribution of phase on the modulator (Fig. 4). Using variance as a measure, we find that $\sigma_{SLM}^2 = E[(X - \mu)^2] = A\alpha^2 \exp[-B\alpha^C]$, where X represents the pixel value on the SLM $\mu = E[X]$ is the mean value of all the pixels, and α is the modulation power. This stretched-exponential form suggests a nontrivial (i.e. non-diffusive) optical flow as the object modes are mixed by the modulator [22, 24-26]. As shown in Fig. 3A-3I, the optimal system response ranges from a strict proportionality to amplitude ($\phi_{SLM} = \gamma\sqrt{I_U}$) for the coherent case to a more conventional Kerr nonlinearity ($\phi_{SLM} = \gamma I_U$) for the incoherent case (consistent with the optical and modulation transfer functions of linear propagation, respectively [21]).

While electro-optic devices have their limitations, they are far more versatile than the physical media traditionally used for nonlinear optics. Of particular significance is the ability to create nonlinearities that are distinct from those of physical media by leveraging independent control of amplitude and phase modulation. This freedom opens new applications for functional forms that have remained purely in the mathematical domain. The ability to adjust parameters dynamically also holds much promise for improving the efficiency and performance of adaptive optics and compressed sensing/imaging. Finally, digital methods have the potential to revolutionize imaging at extremely low light levels, such as fluorescence microscopy and quantum imaging, as SLMs are inherently non-destructive and can operate at the single photon level.

**Acknowledgments:**

This work was funded by the AFOSR.

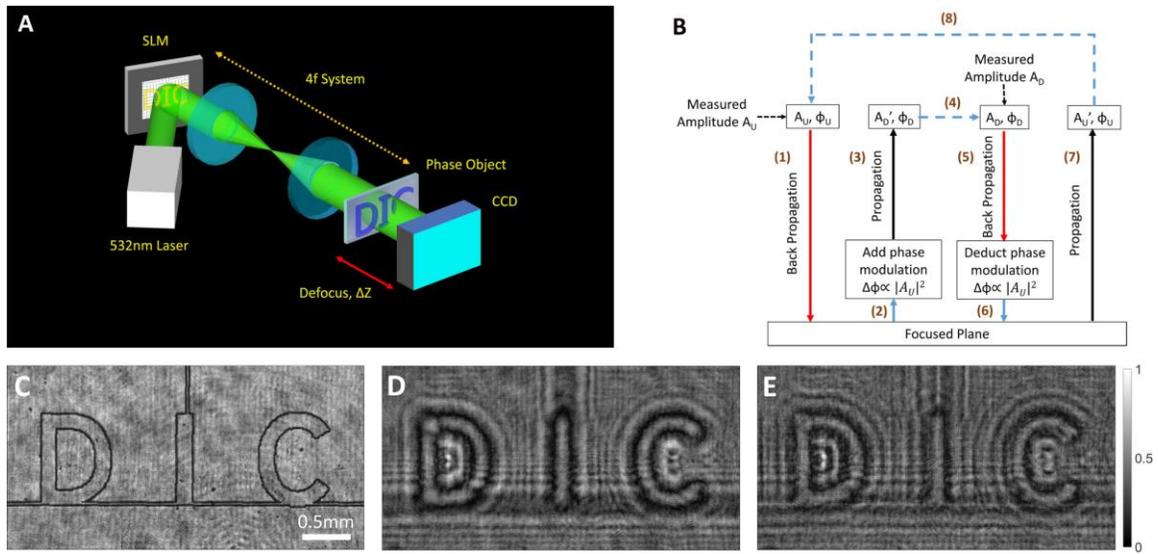

**Fig. 1. Principle of nonlinear digital imaging.** (**A**) Experimental setup. A collimated laser beam is first sent on a reflective SLM. The surface of the SLM is imaged onto the object via a 4-f lens relay. A CCD camera collects the light transmitted through the sample at a defocused distance $\Delta z = 4.5$cm. (**B**) Pseudo-code for the image retrieval algorithm (see text for details). (**C-E**) Experimental intensity measured at (**C**) the focal plane, (**D**) the defocused plane with uniform illumination, and (**E**) the defocused plane with nonlinear (designer) illumination.

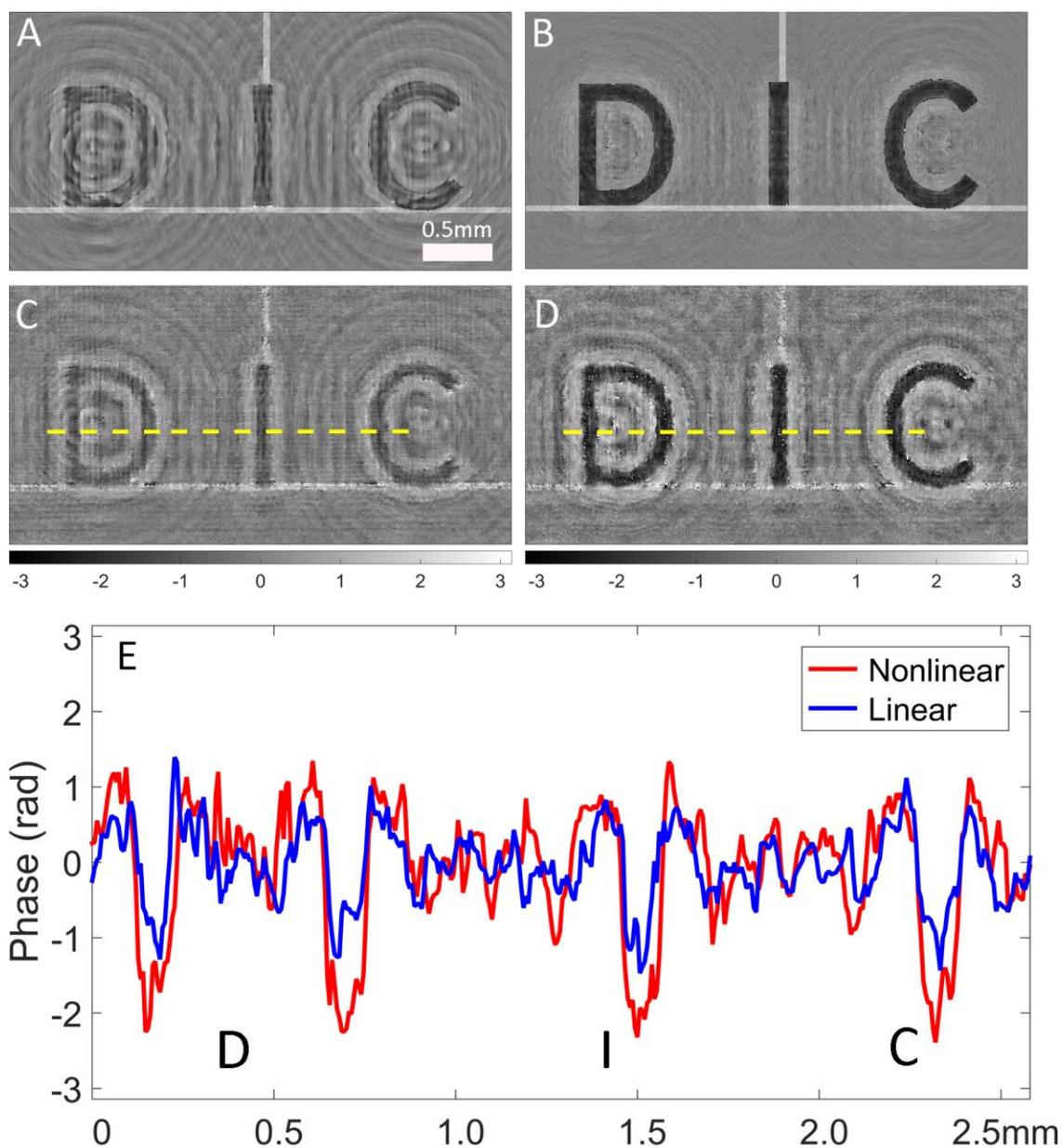

**Fig. 2. Comparison between linear and nonlinear reconstructions.** Simulation of phase reconstruction with (**A**) the linear algorithm and (**B**) the nonlinear algorithm. (**C-D**) Experimental phase reconstruction with (**C**) the linear algorithm and (**D**) the nonlinear algorithm using partially coherent light (corresponding to a speckle spatial frequency of $2.5*10^3$ rad/m). (**E**) Profiles along the dashed lines in (C) and (D).

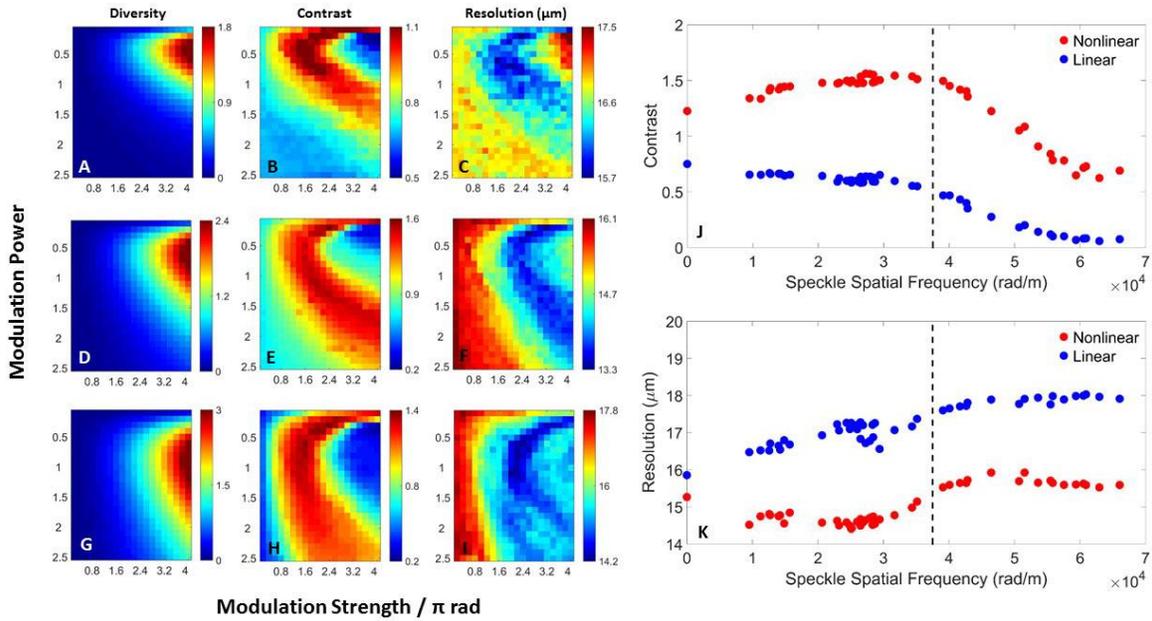

**Fig. 3. Experimental results as a function of nonlinear strength and power.** (A-I) show the phase variance (A,D,G), contrast (B,E,H), and resolution (C,F,I) of the nonlinear reconstructions. (A-C) show the nonlinear reconstructions for coherent illumination while (D-F) and (G-I) are reconstructions for partially coherent illumination with speckle spatial frequency 2.3 rad/m and 4.9 rad/m, respectively. (J,K) Comparison of linear (blue) and nonlinear (red) reconstructions for (J) contrast and (K) resolution. Full coherence is given by zero spatial frequency, and the nonlinearity is fixed at modulation power = 0.5, modulation strength = 1.5π. The dashed line corresponds to the spatial frequency at which the illumination speckle size becomes comparable to the DIC character stroke width.

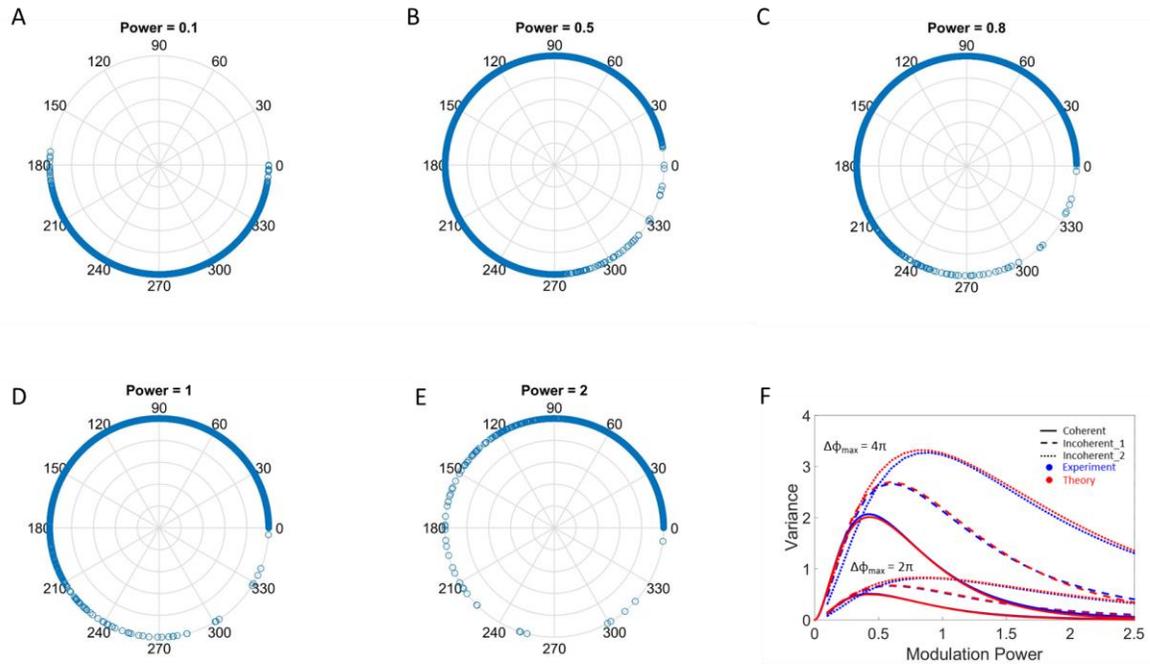

**Fig 4. Diversity of phase modulation. (A-E).** Modulation $\phi_{SLM} = \gamma I^\alpha$ for coherent light (strength $\gamma = 2\pi$). The largest phase/information diversity (for better sampling) occurs when α = 0.5. (**F**) shows the relation between diversity/variance and modulation power. Lower and upper curves represent modulation strengths γ = 2π and 4π, while solid, dashed, and dotted lines denote coherent and partially coherent illumination with speckle spatial frequency 2.3 rad/m (Incoherent_1) and 4.9 rad/m (Incoherent_2). Blue: experiment; red: theoretical fits for stretched exponential $\sigma^2_{SLM} = A\alpha^2 \exp[-B\alpha^C]$, with A = 33 and 132 for the lower and upper curves, B = {4.9, 4.1, 3.7}, and C = {0.8, 0.7, 0.6} for the coherent and incoherent cases.

## *Supplemental Text:*

This Supplemental Information gives more details about the experimental method, the numerical phase-retrieval algorithm, and the digital nonlinear response. It gives an experimental demonstration of digital modulation instability with a phase-only modulator and discusses issues of information diversity and noise.

### 1. Experimental method

A 532nm Coherent Verdi laser was used as a light source, and a phase-only spatial light modulator (SLM, Holoeye PLUTO, with a resolution of 1920 x 1080 pixels, 8-micron pixel size, and 8-bit depth) was placed in the illumination beam before a phase object, modulating the illumination light as a function of the diffracted pattern of the sample. As a direct image does not reveal the object, we defocused the CCD camera (with a resolution of 1280×1024 pixels, 6.7-micron pixel size, and 16-bit depth) a distance $\Delta z$=4.5cm from the image plane and recorded the subsequent diffraction pattern. To produce partially incoherent light, the experimental setup was modified by inserting a rotating diffuser in the illumination path, and the degree of spatial incoherence was quantified by measuring the spatial frequency of the speckles, i.e. $2\pi$ over the correlation length of speckles at the detection plane.
\

### 2. Numerical method

From the two images $I_U$ and $I_D$, the nonlinear phase retrieval algorithm proceeds as follows:

(1) An object field $u_{rec}$ with amplitude $A_U$ and initial uniform phase $\phi = 0$ is numerically back-propagated from the (out-of-focus) camera plane to the sample plane (in focus).
(2) The phase modulation $\phi_{SLM}$ is added to the phase of $u_{rec}$ to simulate the action of the SLM.
(3) The new field is forward-propagated to the camera plane, yielding a field $u'_D$ with amplitude $A'_D$.
(4) The simulated amplitude $A'_D$ is replaced by the amplitude of the actual measurement $A_D$.
(5) The field is back-propagated again to the sample plane, yielding a new field $u_{rec}$.
(6) The phase modulation $\phi_{SLM}$ is subtracted from the phase of $u_{rec}$.
(7) The field is forward-propagated, yielding an estimate of the un-modulated image amplitude $A'_U$.
(8) A new field is generated by replacing $A'_U$ with the actual measurement $A_U$. At this point, the algorithm restarts from step 1.

This algorithm effectively interpolates between the infinitesimal displacement of transport-of-intensity algorithms [27] and the infinite (far-field) displacement of the Gerchberg-Saxton algorithm [14]. That is, it works for any distance of defocus, subject to the limits of Fresnel diffraction (which can be modified by changing the propagator). The method also works for arbitrary degrees of spatial coherence, as in this case the algorithm converges on an overall ensemble phase [28].

### 3. Action of phase-only nonlinearity

**A. Modulation instability (MI)**

Perhaps the simplest nonlinear effect to consider is modulation instability, since its dynamics can be demonstrated easily with a uniform plane wave (matching the uniform illumination used in the experiment). In MI, amplitude perturbations can grow in a self-focusing medium and damp in a self-defocusing one, as intensity-dependent changes in the refractive index create local converging and diverging lens profiles, respectively.

There are two interesting features when modeling MI. The standard theoretical analysis, using linearized perturbation theory, gives a growth rate that is independent of the amplitude of the perturbation. Rather, it represents a competition between nonlinear self-focusing and diffraction, with the former determined by the intensity of the background. In numerical simulation, e.g. using a split-step beam propagation code, the nonlinearity gives a phase change (through the change in refractive index) but the perturbation amplitude in the simulation is retained. Indeed, for propagation (vs. initial growth), it is this amplitude/intensity which is fed back into the code for further evolution of the wave.

As mentioned in the text, the method of digital nonlinear imaging is effectively a physical implementation of the split-step method. For full-field feedback, then, the dynamics should be exactly the same as a traditional beam propagation simulation. For phase-only feedback, the amplitude from the SLM is reset to uniform intensity. It is therefore a hybrid nonlinearity, combining aspects of initial growth with phase-dependent feedback. For a single application of the method, as in the text, this distinction is less significant than a compounded effect from repeated iterations. Indeed, we show in Fig. S1 that the current experimental method reproduces the appropriate features of MI, viz. growth for self-focusing and damping for self-defocusing nonlinearity.

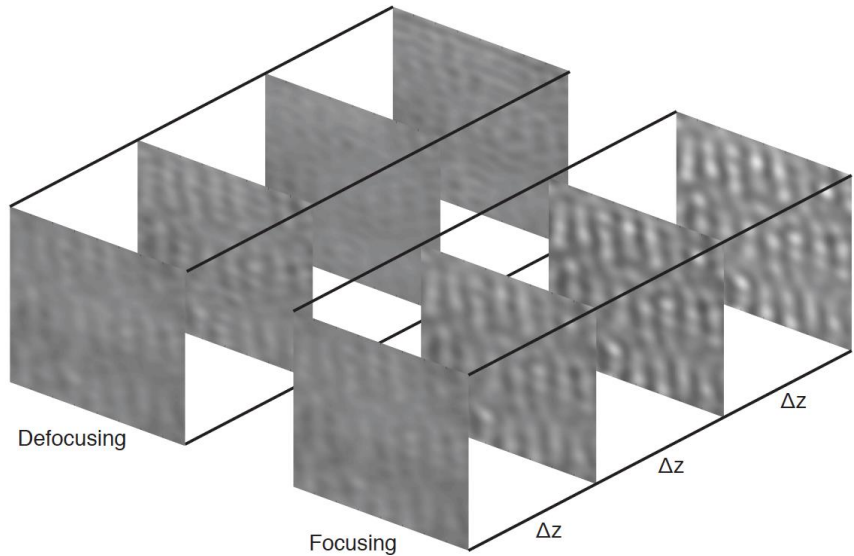

**Fig. S1. Experimental demonstration of modulation instability.** Shown are CCD images after repeated application of a phase-only digital nonlinearity for an initial sinusoidal phase perturbation. As in the usual full-field propagation of MI, the self-focusing case is unstable and the self-defocusing case is stable.

### B. Nonlinear response

Even in the ideal case, with no noise, there is an optimal value of nonlinearity for a given imaging system. For very weak response, there is little difference between linear and nonlinear output, and therefore little gain in information. Stronger responses give more significant differences, evidenced by more pronounced intensity fringes as modes interfere, until high-intensity regions start dominating the image. Above this point, the growth of hot spots leads to a modulation that is not indicative of the object as a whole (an effect exacerbated by noise, e.g. through MI). Examples of this sequence, for coherent and partially coherent light, are shown in Figs. S2, S3.

As discussed in the text, the optimal system response is given by the most diverse distribution of phase on the modulator. This corresponds to the most robust fringes on the feedback pattern (Figs. S3A and S3F).

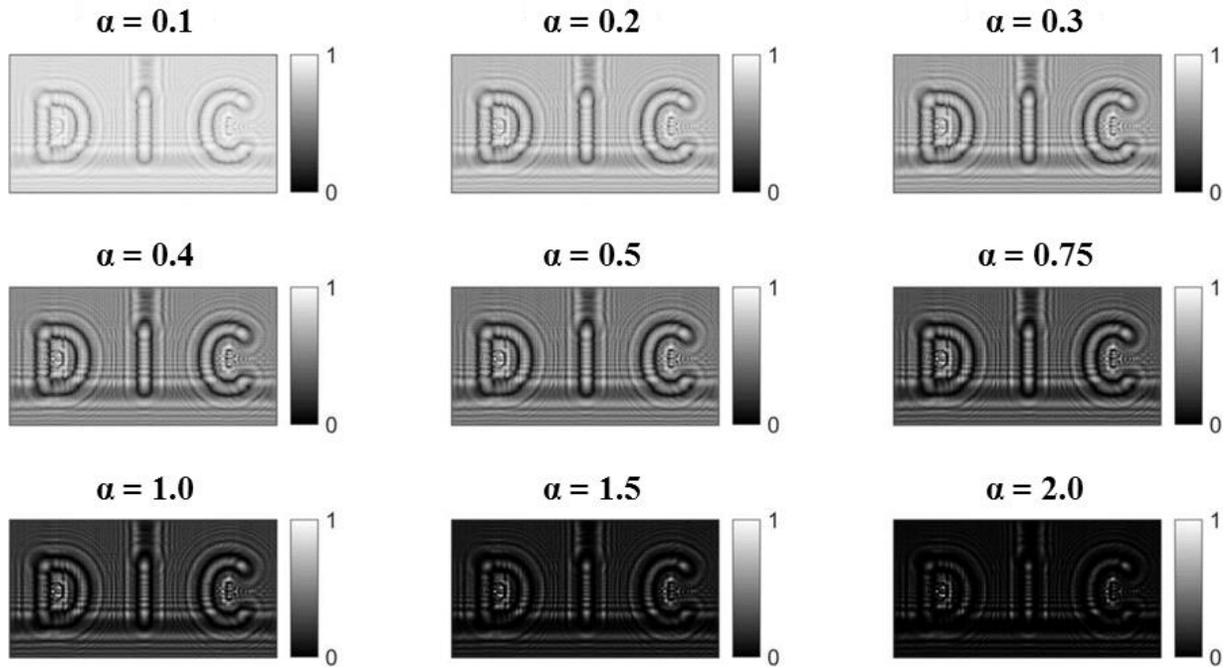

**Fig. S2. Simulated behavior of coherent $I_U^\alpha$.** As the exponent α differs from 0.5, $I_U^\alpha$ becomes more uniform and leads to less phase modulation (intensity fringes).

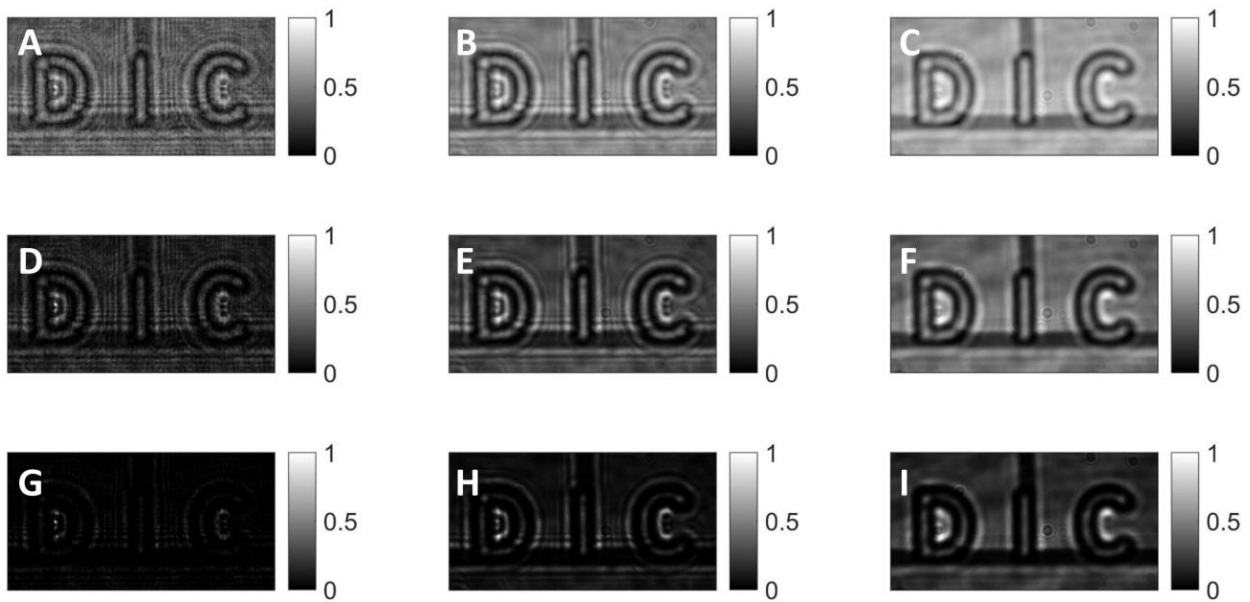

**Fig. S3. Experimental measurements of intensity for different power of modulator response.** Distributions of $I_U^{0.5}$ using (**A**) coherent light, and (**B,C**) partially coherent with speckle spatial frequency (**B**) 2.3 rad/m and (**C**) 4.9 rad/m. (**D-F**) Distributions of $I_U$ corresponding to the same spatial coherence of (A-C). (**G-I**) Distributions of $I_U^2$ corresponding to the same spatial coherence of (A-C).

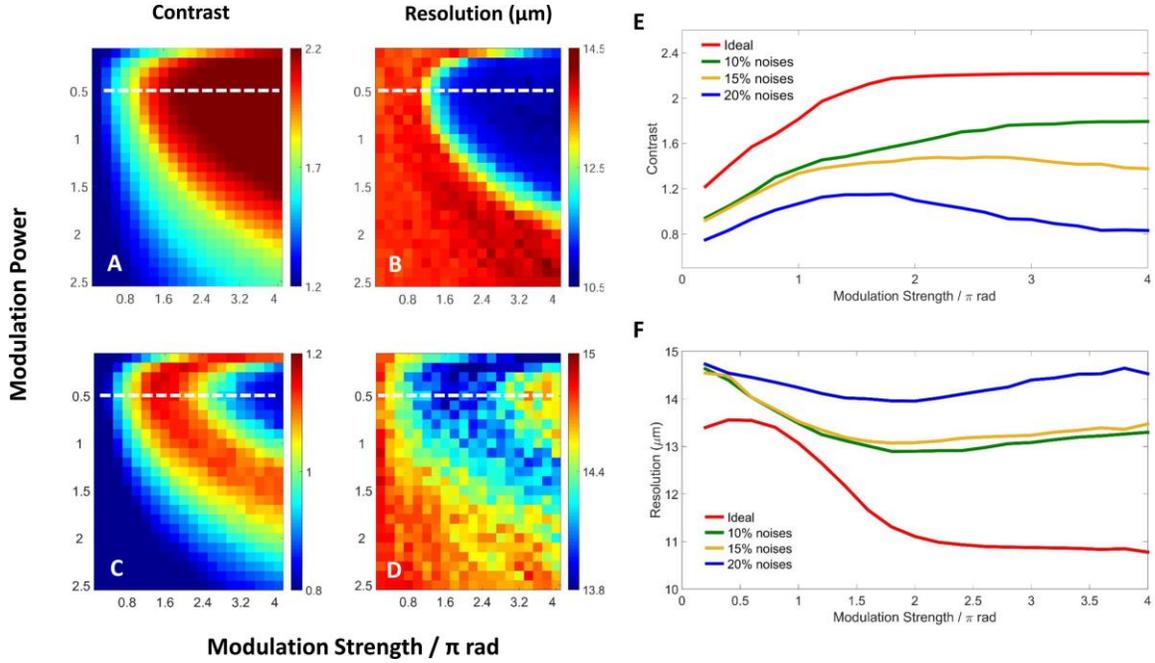

**Fig. S4. Simulation results of system performance with and without noise.** (**A-D**) Contrast and resolution (**A,B**) without and (**C,D**) with noise. (**E,F**) Cross-sections of (A-D) showing competition between nonlinear modulation strength and noise.

## C. Limitations

In practice, the possible nonlinear improvement in image quality is capped by limitations in the camera and/or SLM. Of these, the finite dynamic range is the most significant [23], as the new modes generated by the nonlinearity may be weaker than the noise/intensity floor that can be detected. A numerical simulation of this is shown in Fig. S4.

We note that in the typical spirit of computational optics, many of the single-shot limits in resolution, contrast, and dynamic range can be overcome by taking multiple images over a broader range of scales.